\UseRawInputEncoding
\documentclass[twocolumn]{aastex63}
\shorttitle{Flare-associated CME Candidates on Two M-dwarfs}
\shortauthors{Wang et al.}

\begin{document}

\title{Detection of Flare-associated CME Candidates on Two M-dwarfs by GWAC and Fast, Time-resolved Spectroscopic Follow-ups}

\correspondingauthor{J. Wang, H. L. Li}
\email{wj@nao.cas.cn, lhl@nao.cas.cn}

\author{J. Wang}
\affiliation{Guangxi Key Laboratory for Relativistic Astrophysics, School of Physical Science and Technology, Guangxi University, Nanning 530004, Peopleʼs Republic of China}
\affiliation{Key Laboratory of Space Astronomy and Technology, National Astronomical Observatories, Chinese Academy of Sciences, Beijing 100101,
Peopleʼs Republic of China}

\author{L. P. Xin}
\affiliation{Key Laboratory of Space Astronomy and Technology, National Astronomical Observatories, Chinese Academy of Sciences, Beijing 100101, Peopleʼs Republic of China}

\author{H. L. Li}
\affiliation{Key Laboratory of Space Astronomy and Technology, National Astronomical Observatories, Chinese Academy of Sciences, Beijing 100101, Peopleʼs Republic of China}

\author{G. W. Li}
\affiliation{Key Laboratory of Space Astronomy and Technology, National Astronomical Observatories, Chinese Academy of Sciences, Beijing 100101, Peopleʼs Republic of China}

\author{S. S. Sun}
\affiliation{Guangxi Key Laboratory for Relativistic Astrophysics, School of Physical Science and Technology, Guangxi University, Nanning 530004, Peopleʼs Republic of China}
\affiliation{Key Laboratory of Space Astronomy and Technology, National Astronomical Observatories, Chinese Academy of Sciences, Beijing 100101, Peopleʼs Republic of China}
\affiliation{School of Astronomy and Space Science, University of Chinese Academy of Sciences, Beijing, Peopleʼs Republic of China}

\author{C. Gao}
\affiliation{Guangxi Key Laboratory for Relativistic Astrophysics, School of Physical Science and Technology, Guangxi University, Nanning 530004, Peopleʼs Republic of China}
\affiliation{Key Laboratory of Space Astronomy and Technology, National Astronomical Observatories, Chinese Academy of Sciences, Beijing 100101, Peopleʼs Republic of China}
\affiliation{School of Astronomy and Space Science, University of Chinese Academy of Sciences, Beijing, Peopleʼs Republic of China}

\author{X. H. Han}
\affiliation{Key Laboratory of Space Astronomy and Technology, National Astronomical Observatories, Chinese Academy of Sciences, Beijing 100101, Peopleʼs Republic of China}

\author{Z. G. Dai}
\affiliation{School of Astronomy and Space Science, Nanjing University, Nanjing, 210000, People’s Republic of China}

\author{E. W. Liang}
\affiliation{Guangxi Key Laboratory for Relativistic Astrophysics, School of Physical Science and Technology, Guangxi University, Nanning 530004, Peopleʼs Republic of China}


\author{X. Y. Wang}
\affiliation{School of Astronomy and Space Science, Nanjing University, Nanjing, 210000, People’s Republic of China}


\author{J. Y. Wei}
\affiliation{Key Laboratory of Space Astronomy and Technology, National Astronomical Observatories, Chinese Academy of Sciences, Beijing 100101,
Peopleʼs Republic of China}
\affiliation{School of Astronomy and Space Science, University of Chinese Academy of Sciences, Beijing, Peopleʼs Republic of China}




\begin{abstract}
The flare-associated stellar coronal mass ejection (CME) in solar-like and late type stars is quite essential for the habitability of an exoplanet. 
In this paper, we report detection of flare-associated CMEs in two M-dwarfs, thanks to the high cadence survey carried out by the 
Ground Wide-angle Camera system and the fast photometric and spectroscopic follow-ups. 
The flare energy in $R-$band is determined to be $1.6\times10^{35}\ \mathrm{erg}$ and $8.1\times10^{33}\ \mathrm{erg}$ based on the modeling of their light curves.
The time-resolved spectroscopyic observations start at about 20 and 40 minutes after the trigger in both cases. 
The large projected maximum velocity of $\sim500-700\ \mathrm{km\ s^{-1}}$ suggests that the high velocity wing of their 
H$\alpha$ emission lines are most likely resulted from a CME event in both stars, after excluding the possibility of chromospheric evaporation
and coronal rain. The masses of the CMEs are estimated to be $1.5-4.5\times10^{19}\ \mathrm{g}$ and $7.1\times10^{18}\ \mathrm{g}$.

\end{abstract}



\section{Introduction} \label{sec:intro}

Highly energetic flares with a total energy of $10^{33-39}\ \mathrm{erg}$ have been frequently reported 
for solar-like and late-type main-sequence stars in multi wavelength (e.g., 
Pettersen 1989; Schmitt 1994; Osten et al. 2004; 2005; Huenemoerder et al. 2010; Balona 2015; Notsu et al. 2016;
Van Doorsselaere et al. 2017; Kowalski et al. 2013;
Davenport et al. 2016; Chang et al. 2018; Paudel et al. 2018; Schmidt et al. 2019; Xin et al. 2021). 
It is now commonly accepted that the flares are resulted from the stellar magnetic activity, such as
magnetic reconnection (e.g., Noyes et al. 1984; Wright et al. 2011; Shulyak et al. 2017). In the flares, 
the heated chromospheric plasma moves upward and is confined in the erupted magnetic field line, which leads to 
an expectation of a large-scale expulsion of the confined plasma, i.e., coronal mass ejection (CME),
triggered by magnetic reconnection, if the field eruption is strong enough and the overlying fields are not too constraining
(see a review in Forbes et al. 2006).  The energy released in reconnection
heats the plasma and eject it into interplanetary space (e.g., Tsuneta  1996; Kliem et al. 2000; Karlicky \& Barta 2007; Li et al. 2016).  

The linkage between flares and CMEs has been, in fact, firmly demonstrated in the Sun. Solar observations indicate 
that the CME occurrence, mass, and energy increase with the flare energy\footnote{The relationship is in fact 
roughly valid for stellar CMEs, when their flare X-ray luminosity is involved (Moschou et al. 2019). } (e.g., Yashiro et al. 2008; Aarnio et al. 2011;
Webb \& Howard 2012). However, the CMEs detected on other active stars with stronger magnetic fields and
more energetic flares are still rare at the current stage (e.g., Moschou et al. 2019). The detection and study
of stellar CMEs are essential for evaluating the habitability of an exoplanet, because gas-dynamic simulation
suggests that the frequent stellar CMEs can tear off most of the atmosphere of an exoplanet (e.g., Cherenkov et al. 2017).

Due to a lack of enough spatial resolution, detection of stellar CMEs is still a hard task for current instruments. 
Even though, two methods based on Doppler-shift and X-ray (and EUV) absorption have been developed to detect stellar CMEs. 
We refer the readers to Moschou et al. (2019) for a brief review.
Based on the large blue wing of H$\gamma$ emission line, Houdebine et al. (1990) reported the fastest CME with
a projected maximum velocity of $5800\ \mathrm{km s^{-1}}$ on 
M-dwarf AD Lec. Koller et al. (2021, and references therein) recently identified six CME candidates with 
an excess at the Balmer line wing by a systematic search in the archival SDSS spectral database. 
A more definite evidence of stellar CMEs can be learned from Argiroffi et al. (2019), who detected a delayed and blueshifted 
($-90\pm30\ \mathrm{km\ s^{-1}}$)
\ion{O}{8}$\lambda18.97$\AA\ emission line on active star HR\,9024 by time-resolved high-resolution X-ray spectroscopy 
taken by the Chandra X-ray Observatory space telescope.  
Moschou et al. (2017) reported a temporal decay of the X-ray absorption in the superflare on the eclipsing 
binary Algol, which can be interpreted by an absorption by an expanding CME. A claim of stellar CME based on the 
EUV dimming (e.g., Chandra et al. 2016) can be found in Ambruster et al. (1989).

In this paper, we report a claim of detection of CMEs in the flares of two M-dwarfs triggered 
by the GWAC system. The fast and time-resolved spectroscopy enables us to investigate the 
temporal evolution of the Balmer emission-line 
profiles, and to identify the flare-associated CMEs according to their high-velocity line wings. 
The paper is organized as follows. Section 2 describes the 
discovery of the two flares. The photometry and spectroscopic follow-ups, along with the 
corresponding data reductions, are outlined in Sections 3 and 4, respectively.
Section 5 presents the light curve and spectral analysis. The results and discussion is shown in 
Section 6.

\section{Detection of Flares by GWAC} \label{sec:style}
Ground-based Wide-Angle Cameras (GWAC) system, which aims to detect fast optical transients by a high cadence down 
to 15 seconds, is one of the ground facilities of SVOM mission\footnote{SVOM is a 
China–France satellite mission dedicated to detect and
study Gamma-ray bursts (GRBs). Please see the White Paper given in Wei et al. (2016) for the details.}.
The whole system contains a set of cameras each with a diameter of 18cm and a Field-of-View of 150$\mathrm{deg^2}$, ans a set of follow-up telescopes.
With the cadence, the limiting magnitude of the cameras is down to $R\sim$16.0 mag. 
The follow-up observations in image can be carried out by the dedicated two 60 cm telescopes (GWAC-F60A/B) and
one 30 cm telescope (GWAC-F30), which are all deployed beside the cameras. Now, the GWAC system is located at Xinglong observatory, National Astronomical Observatories of Chinese Academy of Sciences (NAOC).  
We refer the readers to Wang et al. (2020) and Xin et al. (2021) for the more detailed description of the GWAC system.

Table 1 tabulates the log of the two flares, i.e., GWAC\,201221A and GWAC\,211117A, discovered by the GWAC system.
The typical localization error determined from the GWAC images is about 2\arcsec\ in both
cases. Figure 1 shows the discovery images and the corresponding reference images taken by the GWAC cameras. 
In both cases, the shapes of the image profile of the two transients are
very similar to the point spread function (PSF) of the nearby
bright objects, suggesting a high probability that they are not hot pixels. 
The two transients did not show any apparent motion among the several consecutive images. There are also
no known minor planets or comets\footnote{https://minorplanetcenter.net/cgi-bin/mpcheck.cgi?} 
brighter than $V=20.0$mag within a radius of 15\arcmin. Also,
no known variable stars or CVs can be found in SIMBAD around the transient positions within 1\arcmin.

\begin{figure}[ht!]
\plotone{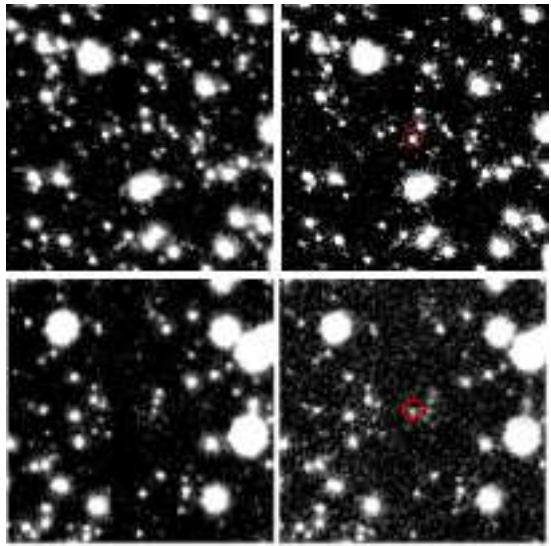}
 \caption{Discovery of GWAC\,201221A (the two upper panels) and GWAC\,210117A (the two lower panels) by the GWAC system. In each row, the left and right panels show the reference 
 image and
 discovery image, respectively. The size of all images is $19.5\arcmin\times19.5\arcmin$. The transients are marked by the red circles in 
 the discovery images.
 \label{fig:general}}
\end{figure}

For each of the cases,
an off-line pipeline involving a standard aperture photometry was performed at the location of the transient and for
several nearby bright reference stars by using the IRAF\footnote{IRAF is distributed by the National Optical 
Astronomical Observatories,
which are operated by the Association of Universities for Research in
Astronomy, Inc., under cooperative agreement with the National Science
Foundation.} APPHOT package, including the corrections of bias, dark
and flat-field and a differential photometry in a standard manner. 
The actual brightness of each transient was then obtained by a calibration by using the SDSS catalogues through 
the Lupton (2005) transformation\footnote{http://www.sdss.org/dr6/algorithms/sdssUBVRITransform.html/ \#Lupton2005}.
The calculated discovery brightness, along with the quiescent brightness, in $R$-band are shown in Column (5) in Table 1.
The corresponding specific flux in $R$-band at the quiescent state is  $f_R=8.8\times10^{-17}\ \mathrm{erg\ s^{-1}\ cm^{-2}}\ \AA^{-1}$ and 
 $1.4\times10^{-16}\ \mathrm{erg\ s^{-1}\ cm^{-2}}\ \AA^{-1}$ for GWAC\,201221A and GWAC\,210117A, respectively.
The distance listed in the Column (10) is 
obtained by an inverse of the parallax provided in the Gaia catalog. With the distance, the total luminosity in $R$-band at the quiescent state
is estimated to be $L_R=1.5\times10^{30}\ \mathrm{erg\ s^{-1}}$ and  $9.4\times10^{29}\ \mathrm{erg\ s^{-1}}$ for  GWAC\,201221A and GWAC\,210117A, respectively.

%

\begin{longrotatetable}
\begin{deluxetable*}{cccccccccc}
\tablecaption{Two Optical Transients Discovered by the GWAC System.}
\tablewidth{700pt}
\tabletypesize{\scriptsize}
\tablehead{
\colhead{Transient ID} & \colhead{Trigger time} & \colhead{R.A.} & \colhead{Decl} & \colhead{Discovery/quiescent mag} &
\colhead{Quiescent counterpart} & \colhead{Gaia DR2 ID} &
\colhead{$G$-band} & \colhead{GBp-GRp} & \colhead{$d$}\\
 & \colhead{(UT)} & \colhead{(J2000)} & \colhead{(J2000)} & \colhead{(mag)} &  &  & \colhead{(mag)} & \colhead{(mag)}  & \colhead{(pc)}\\
}
\colnumbers
\startdata
GWAC\,201221A & 17:50:36 & 07:39:08.8 &  +17:14:32 &  13.7/18.25 & SDSS\,J073908.7+171435.3 & 3169102436090768768 & $17.557\pm0.002$ & 3.01 &  $295.5\pm15.8$ \\
GWAC\,210117A & 16:06:33&  08:39:42.0  &  +20:17:45 &  14.5/17.75 & EPIC\,212002525\dotfill & 664439416447387392 & $17.081\pm0.002$ & 2.98 &  $188.9\pm4.8$ \\
 \enddata
 \tablecomments{Column (1): the ID of the confirmed transient triggered by the GWAC system. Column (2):  the discovery time in UT.  
 Columns (3) and (4): the celestial coordinate at J2000 equinox. Column (5): discovery brightness in $R$-band. Columns (6) and (7): 
 the name of the quiescent counterpart and the corresponding source ID in the Gaia DR2 catalog. Column (8): the quiescent  brightness in $G$-band extracted from 
 the Gaia DR2 catalog. Column (9): the Gaia GBp-GRp color index of the quiescent counterpart. Column (10): the distance in parsec that is obtained by an 
 inverse of the observed parallax. }
\end{deluxetable*}
\end{longrotatetable}

\section{Photometric Follow-ups and Data Reduction}

Each transient was followed-up in photometry immediately by the GWAC-F60A telescope in the standard
Johnson-Cousins $R$-band. The dedicated real-time automatic transient validation system (RAVS, Xu et al. 2020)
enables us to identify the transient in minutes and to carry out a monitor with an adaptive sampling. 
The sampling is optimized basing upon the brightness and the evolution trend of individual target. 

The properties of the quiescent counterparts of the two transients are tabulated in Table 1 as well. Column (7), (8) and (9), respectively, list the $G$-band brightness, GBp-GRp color and distance of the quiescent counterparts, extracted from 
the Gaia DR2 catalog (Gaia Collaboration et al. 2018a).

Raw images taken by the GWAC-F60A telescope were reduced again  by following the standard routine in the IRAF package, 
including bias and flat-field corrections. The standard aperture photometry and calibration based on the SDSS catalogues through 
the Lupton (2005) transformation were adopted for building the light curves.  The final light curves are shown in Figure 2 for the two flares. Note that  
the effect of reddening can be safely ignored in both cases throughout the current study, because the extinctions in the Galactic plane along the line-of-sight are as low as 
$E(B-V)=0.03$mag and $E(B-V)=0.05$mag for GWAC\,201221A and GWAC\,210117A, respectively, based on the updated dust reddening map provided by Schlafly \& Finkbeiner (2011).

\begin{figure}[ht!]
\plotone{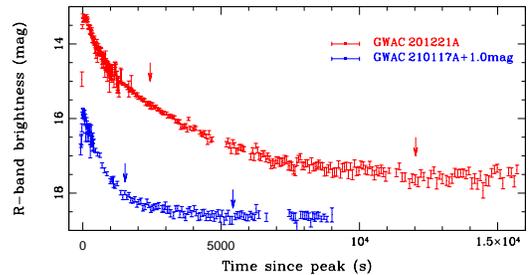}
\caption{$R$-band light curves of GWAC\,201221A (red symbols) and GWAC\,210117A (blue symbols)
observed by GWAC and F60A. The peak times correspond to MJD=59205.243714 day and 
MJD=59232.170581 day for GWAC\,201221A and GWAC\,210117A, respectively. 
For each flare, the arrows mark the start and end of our time-resolved spectroscopy. 
\label{fig:general}}
\end{figure}

\section{Spectroscopy and Data Reduction}

After the discovery and identification of the two flares,  time-resolved long-slit spectroscopy was
performed by the NAOC 2.16m telescope (Fan et al. 2016) as soon as possible through the Target-of-Opportunity mode. 
The log of the spectroscopies are shown in Table 2, where $\Delta t$ is the time delay between the start
of the first exposure of spectroscopy and 
the trigger time. In total, we have 11 and 7 spectra for GWAC\,201221A and GWAC\,210117A, respectively.  The epochs of the start and end of the spectroscopy are marked by the arrows in Figure 2. 

All the spectra were taken by the Beijing Faint Object Spectrograph and Camera (BFOSC) that is 
equipped with a back-illuminated E2V55-30 AIMO CCD. 
The grating G8 with a wavelength coverage  of 5800–8200\AA\ was used in the observations since we focus on
H$\alpha$ emission line. 
With a slit width of 1.8\arcsec\ oriented in the
south–north direction, the spectral resolution is ∼3.5 Å as measured from the sky lines, which corresponds to 
a velocity of $160\ \mathrm{km\ s^{-1}}$ at the H$\alpha$ emission lines. \rm
The wavelength calibrations were carried out with the iron–argon comparison lamps, 
and the flux calibrations with the observations of the Kitt Peak National Observatory standard
stars (Massey et al. 1988). The airmass ranges from 1.1 to 1.3 during the observations.

We reduced the two-dimensional spectra by the standard procedures,
including bias subtraction and the flat-field correction, by using the IRAF package again. For each of the transients, 
fixed apertures of both object and sky emission were used in the spectral extraction for both object and corresponding standard, which is essential for building the
differential spectra (see Section 4 for the details). All of the extracted one dimensional
spectra were then calibrated in wavelength and in flux by the corresponding comparison lamp and standards.
The accuracy of the wavelength calibration is 0.1\AA\ for both flares as assessed by the sky emission lines. 
The extracted spectra are shown in Figure 3. At first glance, a significant descend with time can be 
learned from both H$\alpha$ and \ion{He}{1}$\lambda$6678 emission lines. 
 
\begin{table}[h!]
\renewcommand{\thetable}{\arabic{table}}
\centering
\footnotesize
\caption{Log of Spectroscopic Observations Carried Out by the NAOC 2.16m Telescope. }
\label{tab:decimal}
\begin{tabular}{cccc}
\tablewidth{0pt}
\hline
\hline
ID & Sp. Number & Exposure time (s)  & S/N of H$\alpha$ \\
(1)  &   (2) & (3) & (4)\\
\hline
GWAC\,201221A & \multicolumn{3}{c}{$\Delta t=$40.55min}\\
\cline{2-4}
                            & 1 &  600  & 56.1\\
                            & 2 &  600 & 48.5 \\
                            & 3 &  600 & 55.9 \\
                            & 4 &  600 & 43.4 \\
                            & 5 &  600 & 35.1 \\
                            & 6 &  600 & 39.5 \\
                            & 7 &  1200 & 45.6\\
                            & 8 &  1200 & 39.8\\
                            & 9 &  1800 & 31.6\\
                            & 10 &  1800 & 27.1\\
                            & 11  &  1800 & 31.1\\                            
\hline
GWAC\,210117A & \multicolumn{3}{c}{$\Delta t=$25.55min}\\
\cline{2-4}
                            & 1 &  300  & 20.0\\
                            & 2 &  300  & 16.4\\
                            & 3 &  300  & 11.3\\
                            & 4 &  600  & 14.4\\
                            & 5 &  600  & 20.1\\
                            & 6 &  600  & 24.4\\
                            & 7 &  1200 & 34.2\\
                            \hline
\end{tabular}
\tablecomments{Column (1): the ID of the confirmed transient triggered by the GWAC system. Column (2):  the number series of spectrum. 
Column (3): the exposure time in unit of second. Column (4): the measured signal-to-noise ratio of the H$\alpha$ emission line (see section 5.2 for the details).}
\end{table}

\begin{figure*}[ht!]
\plotone{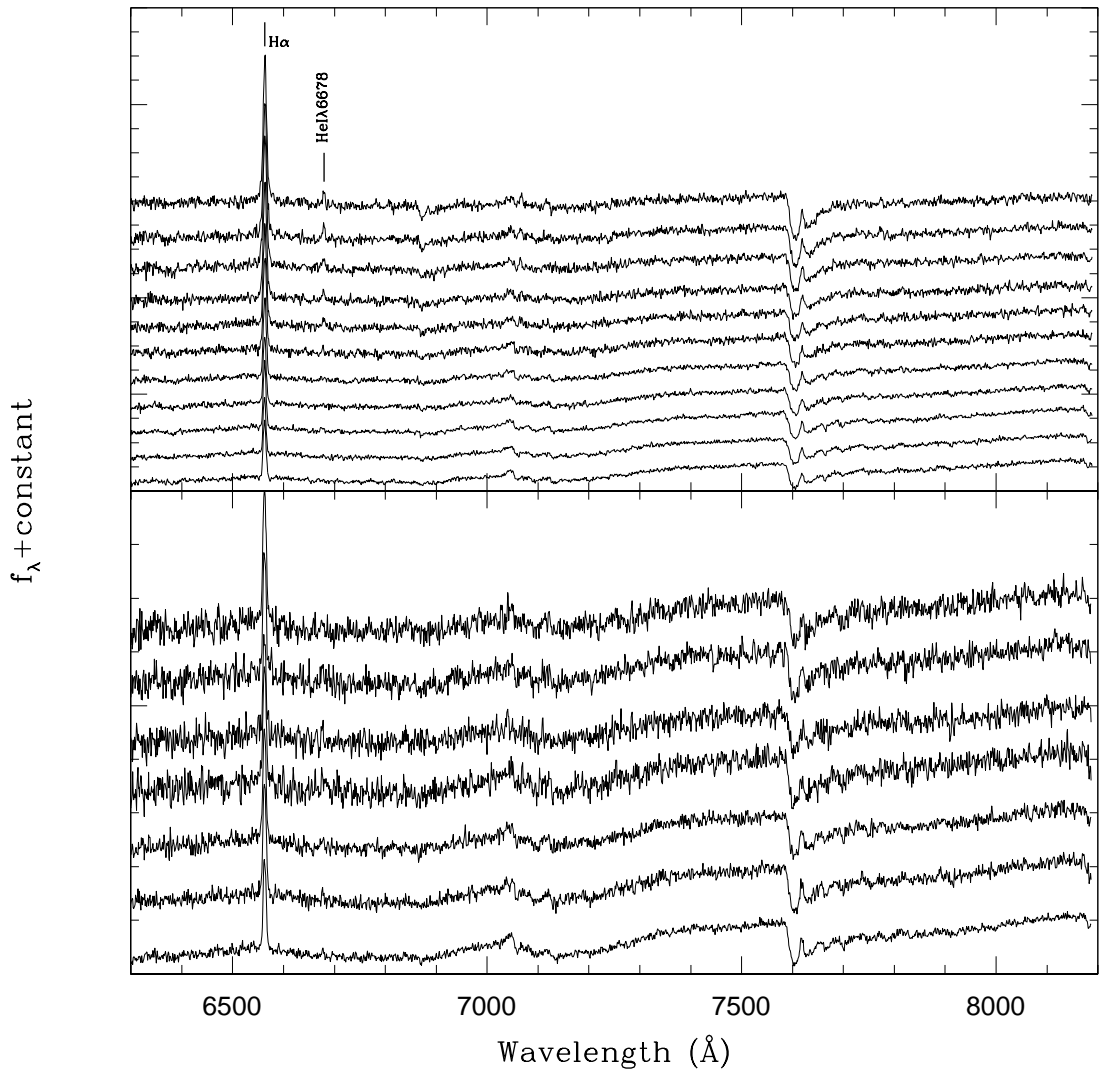}
\caption{
The time-resolved spectroscopies of GWAC\,201221A and GWAC\,210117A are displayed in the upper and lower panels, 
respectively. In each panel, the spectra are sorted with time from top to bottom, and shifted vertically by an arbitrary amount for visibility. The H$\alpha$ and \ion{He}{1}$\lambda$6678 emission lines are marked on the 
top panel.
\label{fig:general}}
\end{figure*}

\section{Light Curve and Spectral Analysis}

\subsection{Light curve analysis}

In order to estimate the total energy released in the flares, the light curves are analyzed by following Xin et al. (2021) who modeled the decay phase of a $\Delta R\sim9.5$ mag flare of a ultra-cool main sequence star more or less by the method described in Davenport et al. (2014).
We at first calculate $F_{\mathrm{amp}}$, the peak relative flux normalized to the quiescent level 
in $R-$band (see Section 2) from the peak brightness, after transforming the GWAC magnitudes to 
the $R-$band through the method described early in Section 2.  
Taking into account of the GWAC's high cadence of 15 second, the detected 
peak magnitude is simply adopted as the real peak brightness of the flare.
$F_{\mathrm{amp}}$ is resulted to be 100 and 20 for GWAC\,201221A and GWAC\,210117A, respectively.

With the determined $F_{\mathrm{amp}}$, we then model each light curve by two separate phases.

\begin{itemize}
 \item \bf Rising phase. \rm By following Xin et al. (2021), we model the rising phase by a linear relationship for the cases with rare detection: 
\begin{equation}
\frac{F_{\mathrm{rise}}}{F_{\mathrm{amp}}}=a_0+k_0t
\end{equation}
 where $F_{\mathrm{rise}}$ is the relative flux normalized to the quiescent level.

\item \bf Decaying phase. \rm After the peak brightness, bot light curves can be modeled by 
a template being composed of a sum of three exponential components,
the template proposed in Davenport et al. (2014). The template used in the current study is a sum of three exponential components, standing for the impulsive decay phase, the gradual decay phase and the late shallow decaying phase: 
\begin{equation}
\frac{F_{\mathrm{decay}}}{F_{\mathrm{amp}}}=\sum_{i=1}^{3}k_ie^{-a_i\frac{t}{t_{\mathrm{1/2}}}}
%
\end{equation}
where $t_{1/2}=1$, a parameter describing the full time width at half the maximum flux,
is adopted in our modeling (Xin et al. 2021).

\end{itemize}

The best modelings of the light curves of the two flares are shown in Figure 4. The best-fit 
parameters, along with the reduced $\chi^2$, are tabulated in Table 3. 
In GWAC\,210117A, a few of measurements with an evident deviation are excluded in the modeling. 
We argue that the corresponding small value of the reduced $\chi^2$ is likely due to the large photometric
uncertainty at the end of the decaying.
With the modeling, 
the equivalent duration (ED) of a flare, defined as the time needed to emit all the flare energy at a quiescent flux level (e.g., Kowalski et al. 2013), is therefore determined to be 
11223 and 8820 seconds for GWAC\,201221A and GWAC\,210117A, respectively.

\begin{figure}[ht!]
\plotone{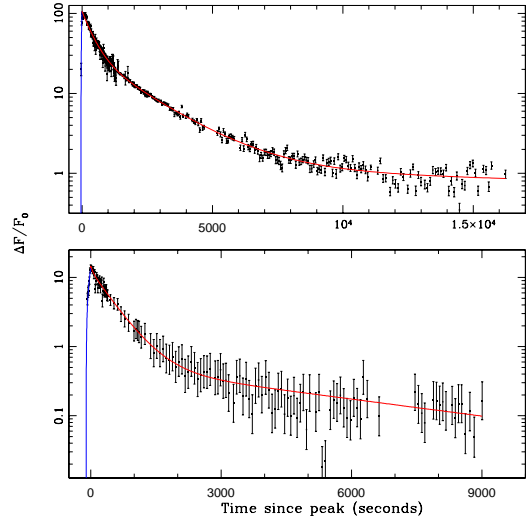}
\caption{Light curves of the relative flare flux of 
GWAC\,201221A (the upper panel) and GWAC\,210117A (the lower panel).
In each panel, the best fitted models in the rising phase and decaying phase are denoted by the blue and red lines, respectively.
\label{fig:general}}
\end{figure}

\begin{table}[h!]
\renewcommand{\thetable}{\arabic{table}}
\centering
\footnotesize
\caption{Best-fit parameters of the light curves of the two flares.}
\label{tab:decimal}
\begin{tabular}{cccc}
\tablewidth{0pt}
\hline
\hline
Parameter & Value & Parameter & Value \\
(1)  &   (2) & (3) & (4) \\
\hline
\multicolumn{4}{c}{GWAC\,201221A ($\chi^2/\mathrm{dof}=1.36$)}\\
\hline
 $a_0$ & $1.060\pm0.039$    & $k_0$ & $0.027\pm0.002$   \\
 $a_1$ & $0.00002\pm0.00002$    & $k_1$ & $0.012\pm0.002$ \\
 $a_2$ & $0.00249\pm0.00007$    & $k_2$ & $0.075\pm0.013$ \\
 $a_3$ & $(5.1\pm0.2)\times10^{4}$    & $k_3$ & $0.307\pm0.001 $ \\
\hline
\multicolumn{4}{c}{GWAC\,210117A ($\chi^2/\mathrm{dof}=0.25$)}\\
\hline
 $a_0$ & $0.6896\pm0.04443$ & $k_0$ & $0.007\pm0.001$   \\
 $a_1$ & $0.0021\pm0.0004$  & $k_1$ & $0.595\pm0.065$  \\
 $a_2$ & $0.0187\pm0.0280$  & $k_2$ & $0.105\pm0.082$ \\
 $a_3$ & $0.0002\pm0.0004$  & $k_3$ & $0.027\pm0.040$ \\
\hline
\end{tabular}
\end{table}

\subsection{Spectral analysis}

In this section, we perform spectral analysis on the two flares by focusing on the profiles of the H$\alpha$ emission lines.
The analysis enables us to identify a (redshifted) broad component that is potentially contributed by a  
flare-associated CME. 

The signal-to-noise (S/N)
ratio\footnote{In the estimate of S/N ratio of an emission line, the statistic error of the line $\sigma_l$ is determined by the 
method given in Perez-Montero \& Diaz (2003): $\sigma_l=\sigma_c\sqrt{N[1+\mathrm{EW}/(N\Delta)]}$, where 
$\sigma_c$ is the standard deviation of continuum in a box near the line, $N$ the number of pixels used to measure the line flux, 
$\mathrm{EW}$ the equivalent width of the line, and $\Delta$ the wavelength dispersion in units of $\mathrm{\AA\ pixel^{-1}}$. } of the H$\alpha$ emission line in each spectrum is presented in the Column (4) of Table 2. With the S/N ratios and the profile decomposition described below,
the minimum detectable CME mass $M_{\mathrm{CME,min}}$ could be estimated for the two events 
according to the Equation (9) in Odert et al. (2020), i.e., 

\begin{equation}
  M_{\mathrm{CME,min}}\simeq \frac{\pi R^2_*m_{\mathrm{H}}N_{\mathrm{H}}}{\mathrm{SNR}\times W[1-e^{-\tau}]}
\end{equation}
where $R_*$ is the radius of a host star, $m_{\mathrm{H}}$ the mass of hydrogen atom and $N_{\mathrm{H}}$
the column density of a prominence. $W$ and $\tau$ are the geometric dilution factor and optical depth of 
H$\alpha$ emission line, respectively. 
By adopting the typical value of $W=0.5$, $N_{\mathrm{H}}$ and $\tau=10$ (Odert et al. 2020), $M_{\mathrm{CME,min}}$ is inferred 
to be $2.2\times10^{16}$g and $6.6\times10^{16}$g for GWAC\,201221A and GWAC\,210117A, respectively, when 
the basic properties of the host stars are adopted (see Table 5 in Section 6.1).

\subsubsection{GWAC\,201221A}

In order to examining the variation of the H$\alpha$ line profile, a set of differential spectra are created by subtracting directly the 
last spectrum (i.e., No 11), which is guaranteed by the fixed object and sky apertures used in the extraction.  The differential spectra of 
No. 1-10 are displayed in Figure 5. One can see from the figure that the No. 10  differential spectrum is dominated by random noise,
which suggests a status close to the quiescent level for the last two spectra (i.e., No. 10 and 11, see
also in Figure 2). 

With the differential spectra, 
we then model the H$\alpha$ line profile by the SPECFIT task (Kriss 1994) in the IRAF package, which is 
shown in Figure 6, except the No. 10 spectrum. 
In each differential spectrum, the H$\alpha$ line profile is reproduced by a sum of a linear continuum and 
a set of Gaussian components. Two    
Gaussian components: one is narrow and another is broad ($\mathrm{FWHM}\sim1000\ \mathrm{km\ s^{-1}}$),
are necessary for fitting the profile properly for the No. 1 to 6 spectra. However, 
the profiles in the next three spectra can be modeled properly by 
a single narrow Gaussian function. 

The results of the line profile modeling are tabulated in Table 4. All the reported flux of the H$\alpha$ narrow component 
($f\mathrm{(H\alpha_n)}$) includes the contribution from the No. 11 spectrum of $f_0(\mathrm{H\alpha})=7.9\times10^{-15}\ \mathrm{erg\ s^{-1}\ cm^{-2}}$. Note that a correction of instrumental spectral
resolution is ignored for all the reported widths of the narrow components. In contrast, a correction of 
$\sigma=\sqrt{\sigma_{\mathrm{obs}}^2-\sigma_{\mathrm{inst}}^2}$ is applied for all the broad components, where 
$\sigma_{\mathrm{inst}}$ is the instrumental spectral resolution. The line shifts shown in Columns (6) and (7) are obtained from 
$\Delta\upsilon=c\Delta\lambda/\lambda_0$, where $\lambda_0$ and $\Delta\lambda$ are the rest-frame wavelength
in vacuum of a given emission line and the wavelength shift of the line center, respectively.
Column (8) lists the maximum projected velocity that is measured, from the 
observed spectrum, at the position where the H$\alpha$ high-velocity red wing mergers to the continuum.
Only the uncertainties, which are at the 1$\sigma$ significance level, due to the modeling are reported in the table.

\begin{figure*}[ht!]
\plotone{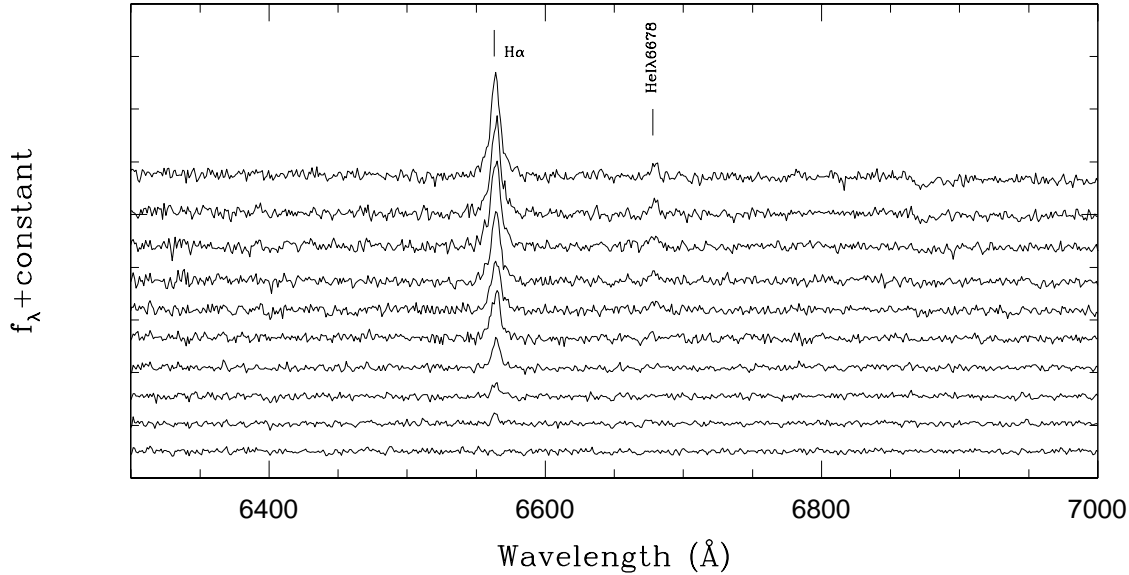}
\caption{The same as in Figure 2 but for the differential spectra of GWAC\,201221A. 
\label{fig:general}}
\end{figure*}

\begin{figure*}[ht!]
\plotone{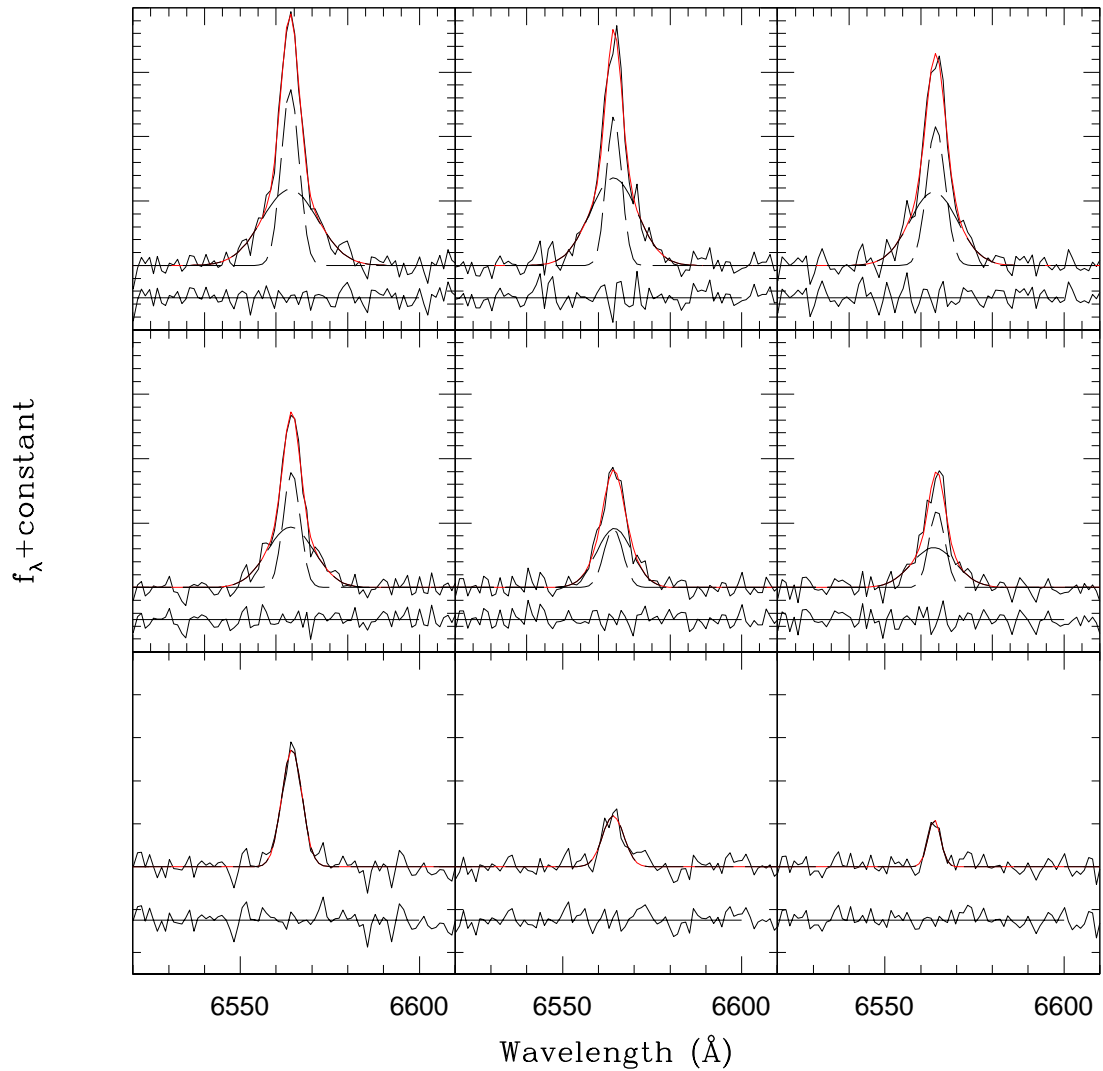}
\caption{An illustration of line profile modelings by a linear combination of a set of Gaussian functions for the H$\alpha$ emission lines in
GWAC\,201221A. In each panel, the modeled local continuum has already been removed from the original observed spectrum. The observed and modeled line profiles are plotted by black and red
solid lines, respectively. Each Gaussian function is shown by a dashed line. The sub-panel underneath the line spectrum presents the residuals between the observed
and modeled profiles.
\label{fig:general}}
\end{figure*}

\begin{table*}[h!]
\renewcommand{\thetable}{\arabic{table}}
\centering
\footnotesize
\caption{Results of Spectral Measurements and Analysis.}
\label{tab:decimal}
\begin{tabular}{cccccccccc}
\tablewidth{0pt}
\hline
\hline
ID & $f\mathrm{(H\alpha_n)}$ & $f\mathrm{(H\alpha_b)}$ & $\lambda_{\mathrm{H\alpha_n}}$  &   $\lambda_{\mathrm{H\alpha_b}}$  & $\mathrm{FWHM(H\alpha_n)}$  &  $\mathrm{FWHM(H\alpha_b)}$  & $\Delta\upsilon(\mathrm{H\alpha_n})$ & $\Delta\upsilon(\mathrm{H\alpha_b})$ & $V_{\mathrm{max}}$\\
 &  \multicolumn{2}{c}{$(\mathrm{10^{-15}erg\ s^{-1}\ cm^{-2}})$} &  \multicolumn{2}{c}{$(\mathrm{\AA})$} & \multicolumn{5}{c}{($\mathrm{km\ s^{-1}})$} \\
(1)  &   (2) & (3) & (4) & (5) & (6) & (7) & (8) & (9) & (10) \\
\hline
\multicolumn{10}{c}{GWAC\,201221A}\\
\hline
1    &  $16.4\pm1.1$  & $11.5\pm1.1$  & $6563.9\pm0.2$ &  $6563.8\pm0.5$  & $270\pm20$ &   $770\pm80$   &  $51\pm9$  &  $47\pm23$ & 720 \\
2   &   $14.5\pm1.2$ & $11.3\pm1.1$  & $6564.3\pm0.2$ &  $6564.2\pm0.4$  &  $230\pm20$ &   $630\pm70$   &  $68\pm9$ &  $64\pm18$ & 610 \\
3   &  $14.9\pm1.1$  & $9.5\pm1.1$   & $6564.3\pm0.2$ &  $6563.6\pm0.5$  & $280\pm20$ &   $630\pm70$   &  $68\pm9$ &  $37\pm23$ & 630\\
4   &  $13.2\pm0.8$ & $7.7\pm0.8$  & $6564.4\pm0.2$ &  $6564.1\pm0.5$  & $250\pm20$ &   $620\pm60$   &  $73\pm9$ &  $59\pm23$ & 570\\
5   &   $10.9\pm2.0$ & $5.7\pm2.0$ & $6564.2\pm0.2$ &  $6564.5\pm0.8$  & $280\pm70$ &   $410\pm80$    &  $64\pm27$  &  $78\pm36$ & 390\\
6   &  $11.2\pm2.2$  & $4.7\pm2.0$  & $6564.5\pm0.7$ &  $6563.6\pm0.8$  &  $240\pm70$ &   $570\pm260$  &  $78\pm32$ &  $37\pm36$  & 520\\
7   &  $11.8\pm0.2$  & \dotfill           & $6564.4\pm0.2$ &  \dotfill                  &  $300\pm20$ &   \dotfill            &  $73\pm9$ &  \dotfill &  \dotfill \\
8   &  $9.7\pm0.2$  & \dotfill            & $6564.1\pm0.4$ &  \dotfill                  &  $320\pm40$ &   \dotfill            &  $59\pm18$  &  \dotfill &  \dotfill \\
9   &  $8.9\pm0.2$  & \dotfill            & $6563.7\pm0.4$ &  \dotfill                  &  $190\pm20$ &   \dotfill             &  $41\pm18$ &  \dotfill &  \dotfill \\
\hline
\multicolumn{10}{c}{GWAC\,210117A}\\
\hline
1    &  $28.2\pm1.9$  & $5.7\pm1.1$ & $6562.7\pm0.3$ &  $6566.9\pm1.9$     &   $400\pm30$ &  $500\pm110$  &  $-49\pm13$  &  $210\pm90$ & $530$ \\
2   &  $17.8\pm1.2$  & \dotfill  &        $6562.6\pm0.2$ &  \dotfill                   &   $360\pm30$ &  \dotfill  &  $-54\pm11$  &  \dotfill & \dotfill \\
3    &  $16.4\pm1.2$  & \dotfill  &       $6562.6\pm0.3$ &  \dotfill                   &  $340\pm30$ &  \dotfill  &  $-56\pm13$  &  \dotfill & \dotfill \\
4   &   $17.5\pm1.1$  & \dotfill  &        $6562.6\pm0.2$ &  \dotfill                    &  $370\pm30$ &  \dotfill  &  $-56\pm11$  &  \dotfill & \dotfill \\
5   &   $16.8\pm0.6$  & \dotfill  &      $6562.4\pm0.1$ &  \dotfill                    & $360\pm10$ &  \dotfill  &  $-64\pm6$  &  \dotfill & \dotfill \\
6    &  $16.8\pm0.7$  & \dotfill  &      $6562.7\pm0.1$ &  \dotfill                    &  $360\pm10$ &  \dotfill  &  $-51\pm6$  &  \dotfill & \dotfill \\
7    &  $12.5\pm0.4$  & \dotfill  &      $6562.9\pm0.1$ &  \dotfill                    & $320\pm10$ &  \dotfill  &  $-42\pm5$  &  \dotfill & \dotfill \\
\hline                           
\end{tabular}
\tablecomments{Column (1): the number of spectrum in time series. Columns (2) and (3):  the total flux in unit of $\mathrm{10^{-15}erg\ s^{-1}\ cm^{-2}}$
of the H$\alpha$ narrow and broad component, respectively.
Each component is denoted by a Gaussian function.
Columns (4) and (5): the central wavelength unit of \AA of the H$\alpha$ narrow and broad component, respectively.
Columns (6) and (7): the line width (full width at half maximum)  unit of $\mathrm{km\ s^{-1}}$ of the H$\alpha$ narrow and broad component, respectively.
Columns (8) and (9): the bulk velocity shift in unit of $\mathrm{km\ s^{-1}}$ with respect to the rest-frame wavelength of H$\alpha$ line. 
Column (10): the maximum velocity in unit of $\mathrm{km\ s^{-1}}$ of the broad H$\alpha$ line red wing (see the main text for the details). } 
\end{table*}

\subsubsection{GWAC\,210117A}
In this flare,
the relatively low S/N ratio of the continuum prevent us from building the differential spectra. 
We instead model the H$\alpha$ emission lines from the normalized spectra by a single Gaussian function. 
The results are tabulated in Table 4 as well.

Figure 7 compares the temporal evolution of the H$\alpha$ line profiles after a proper combination of the spectra taken at different epochs.  
The comparison clearly shows that there is a weak broad component in the first spectrum (the red curve). 
We then model the corresponding profile by a sum of two Gaussian functions as same as we did for GWAC\,201221A to
reproduce the red line wing.
The best fit is shown in the inert panel in Figure 7, and presented in Table 4 as well. 
Again, a correction of the instrumental resolution is applied for the measured line width of the broad component.  

\begin{figure}[ht!]
\plotone{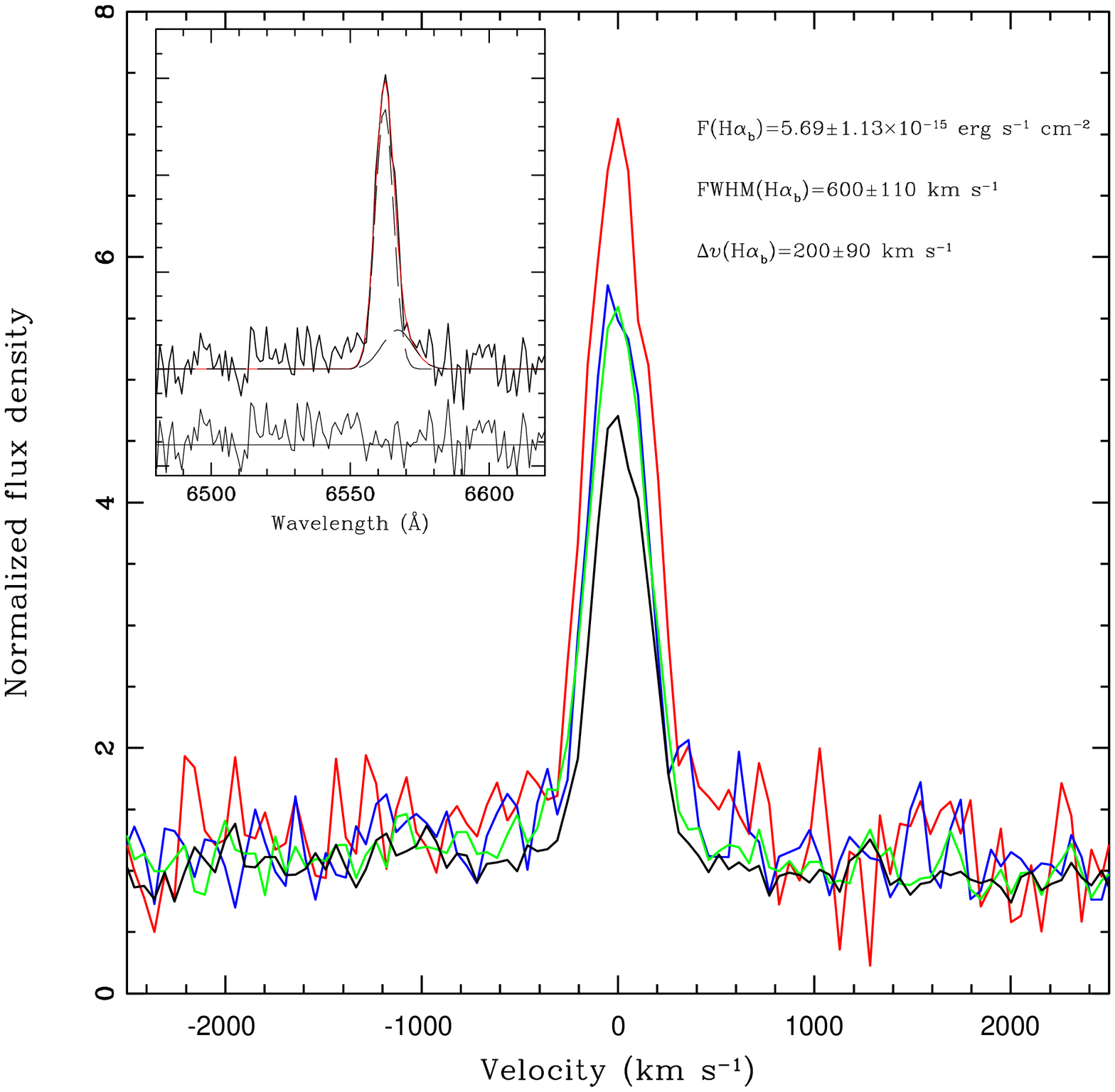}
\caption{A comparison of the H$\alpha$ line profiles of GWAC\,210117A, after a normalization with respect to the 
local continuum. The red and black curves correspond to the first (i.e., No. 1) and last (i.e., No. 7)  spectra, respectively.
The blue curve shows the profile after a combination of No. 2 and 3 spectra, and the green one the profile created from 
a combination of No. 4, 5 and 6 spectra. The insert panel shows the spectral modeling by s sum of two Gaussian profiles
for the first spectrum. The symbols are the same as in Figure 6. \label{fig:general}}
\end{figure}

\section{Results and Discussion}

\subsection{Quiescent Properties of the Two Stars}
Based on their basic properties, we here argue that the host stars of both triggered events are M-dwarfs. 
With their $G$-band absolute magnitudes and $\mathrm{G_{BP}-G_{RP}}$ colors,  
the two host stars are marked on the color-magnitude diagram (CMD) in Figure 8.
One can see from the figure that 1) the host star of GWAC\,210117A follows the 
main sequence of single stars very well; 2) the host star of GWAC\,201221A locates slightly above GWAC\,210117A. This elevation implies that the host star of GWAC\,201221A is 
either a unresolved binary or a young main-sequence star (e.g., Gaia Collaboration et al. 2018b).
 
In addition, some basic parameters quoted from literature are tabulated in Table 5. 
The first row in the table shows the parameters of the host star of GWAC\,201221A determined from photometry by combing Gaia, Pan-STARRS1, 2MASS, and AllWISE by Anders et al. (2019).
In the case, the spectral type of M4 is determined by the recently updated relationship between spectral type and $T_{\mathrm{eff}}$ (e.g., Pecaut \& Mamajek 2013; Stassun et al. 2019).
The parameters of the host star of GWAC\,210117A are extracted from the updated stellar properties for the K2 stars in the Ecliptic Plane Input Catalog. 
The stellar properties, along with the spectral type of M4, are 
built by combining the Gaia distance and spectroscopy of the Large Sky Area Multi-Object Fiber Spectroscopic Telescope (LAMOST) DR5 (Hardegree-Ullman et al. 2019). 
However, a lower $T_{\mathrm{eff}}$ of $\sim3000-3400$K, being more comparable with its spectral type of M4,
can be found in other studies based on photometry  
measurements (e.g., Boudreault et al. 2012; Ilin et al. 2019; Stassun et al. 2019; Cantat-Gaudin et al. 2020). 
In both events, the radius and mass of the host stars are extracted from Stassun et al. (2019).

 \begin{figure}[ht!]
\plotone{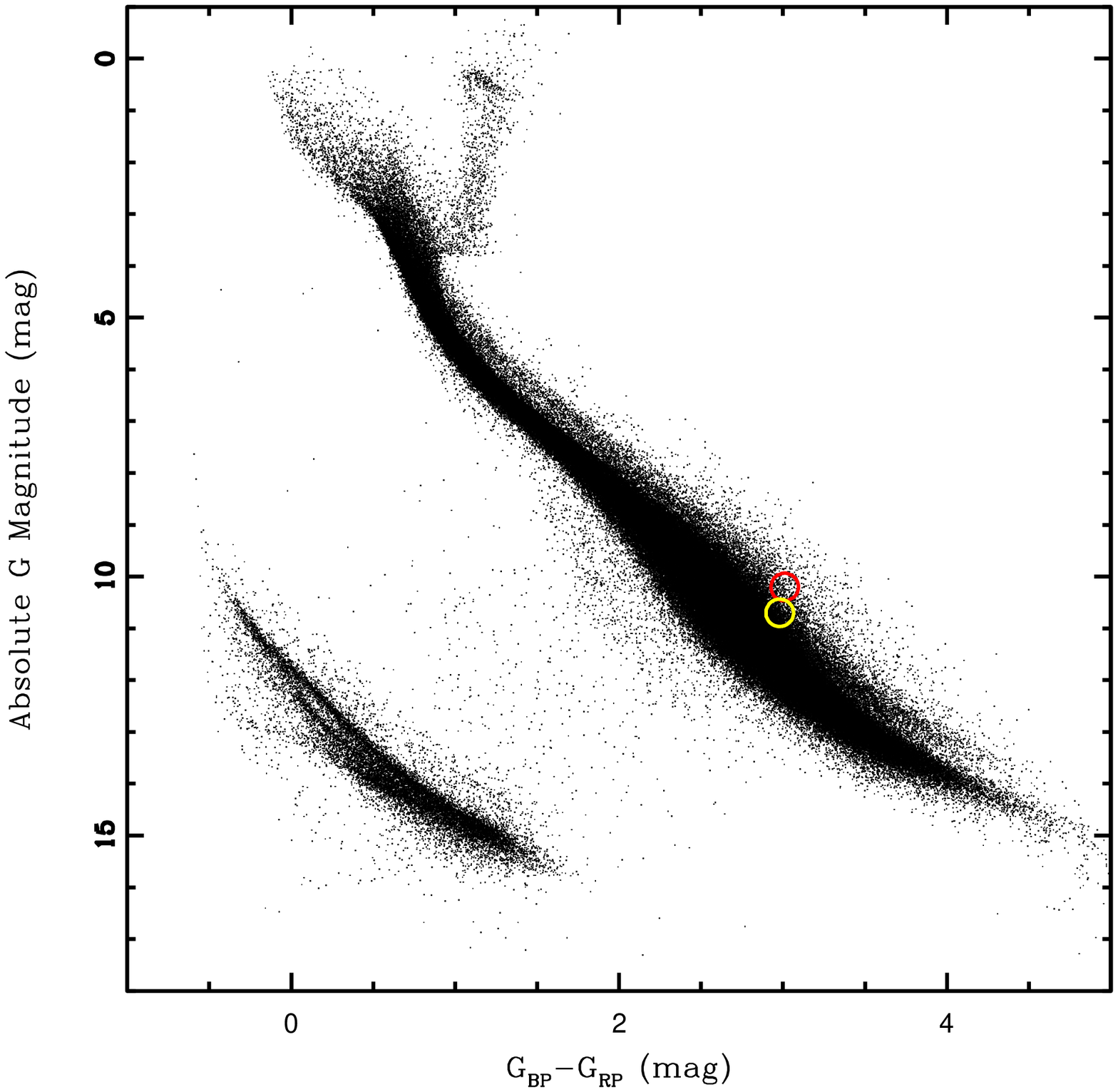}
\caption{CMD of the Gaia stars. The stars at quiescent status are marked by the red and yellow circles for GWAC\,201221A and GWAC\,210117A, respectively.   \label{fig:general}}
\end{figure}

\begin{table*}[h!]
\renewcommand{\thetable}{\arabic{table}}
\centering
\footnotesize
\caption{Basic parameters of the two host stars.}
\label{tab:decimal}
\begin{tabular}{ccccccccc}
\tablewidth{0pt}
\hline
\hline
GWAC ID & Quiescent counterpart & $T_{\mathrm{eff}}$ & $\log g$ &  $\mathrm{Fe/H}$ & Sp. Type & $M_*$ & $R_*$ & References \\
  & &  (K) & ($\mathrm{cm\ s^{-2}}$) & & & ($M_\odot$)  & ($R_\odot$) \\ 
(1)  &   (2) & (3) & (4) & (5) & (6) & (7) & (8) & (9)\\
 \hline
GWAC\,201221A & SDSS\,J073908.7+171435.3 & $3213^{+10}_{-10}$ & $4.780^{+0.005}_{-0.007}$ & $0.491^{+0.005}_{-0.186}$ & M4 & $0.40\pm0.03$ & $0.41\pm0.02$ &  1,2  \\
GWAC\,210117A & EPIC\,212002525\dotfill  & $3694\pm138$ & $4.87\pm0.06$ & $0.114\pm0.235$ & M4 & $0.32\pm0.04$ & $0.34\pm0.01$ &  2,3\\
\hline
\hline                          
\end{tabular}
\tablecomments{Column (1): the ID of the confirmed transient triggered by the GWAC system. Column (2):  the name of the quiescent counterpart. 
Column (3): surface effective temperature in unit of K. Column (4):  logarithmic of surface gravity in unit of $\mathrm{cm\ s^{-2}}$. Column (5): 
Column (5): logarithmic of Fe to H abundance. Column (6): Spectral type. Columns (7) and (8): Mass and radius of the host star in unit of solar mass and solar radius, respectively. Column (9): the reference for the parameters: 1: Andres et al. (2019), 2: Stassun et al. (2019), 2: Hardegree-Ullman et al. (2020).   } 
\end{table*}

\rm
\subsection{Variation of H$\alpha$ Emission Lines}

Based on the spectral analysis described in Section 5.2,
the temporal evolution of the H$\alpha$ emission lines are shown in Figures 9 and 10 for GWAC\,201221A and 
GWAC\,210117A, respectively. As expected in a flare, the total H$\alpha$ line flux fades with time in both events.
In GWAC\,201221A, there is a significant variation of the broad H$\alpha$ component in flux, line width 
and maximum velocity, even though a lack of variation in the bulk velocity.

The measured large maximum velocities ($V_{\mathrm{max}}$) motivate us to argue that the detected broad H$\alpha$ emission is potentially resulted from the flare-associated CMEs in both events. 
At first, even though a blueshift signature leads to a more conclusive CME,
both blueshift and redshift signatures could be observed as argued in Mouschou et al. (2019), 
both because of the projection effect and because of the 
random direction of an ejection (see Figure 4 in Mouschou et al. (2019)).
Secondly, 
with the mass and radius listed in Table 5, the escape velocity 
$\upsilon_{\mathrm{esc}}=630(M/M_\odot)^{1/2}(R/R_\odot)^{-1/2}\ \mathrm{km\ s^{-1}}$
can be determined to be $620\ \mathrm{km\ s^{-1}}$ and $610\ \mathrm{km\ s^{-1}}$
at the stellar surface for GWAC\,201221A and GWAC\,210117A, respectively. In GWAC\,201221A,
the determined $\upsilon_{\mathrm{esc}}$ is slightly smaller than the maximum of the 
measured $V_{\mathrm{max}}$. Although the 
measured $V_{\mathrm{max}}$ is slightly smaller than the $\upsilon_{\mathrm{esc}}$ in GWAC\,210117A, it is noted 
that the 1) the actual ejection velocity might be significantly larger than $V_{\mathrm{max}}$ 
due to the projection effect (e.g., Houdebine et al. 1990), and 2)
a smaller $\upsilon_{\mathrm{esc}}$ is expected if the plasma is ejected above the stellar surface 
(e.g., Lewis \& Simmett 2002) because $\upsilon_{\mathrm{esc}}$ decrease with radius.
\rm

In addition to the CME scenario, the asymmetries of Balmer emission lines could be explained by chromospheric evaporation 
(Canfield et al. 1990; Gunn et al. 1994; Berdyugina et al. 1999), chromospheric condensation 
or coronal rain  (e.g.,  Antolin et al. 2012; Lacatus et al. 2017; Vida et al. 2019; Fuhrmeister et al. 2018). 
In the evaporation scenario, the chromospheric material can move either upward or downward 
due to the injected energy of the flare-accelerated electrons. On the Sun, the typical velocity of the chromospheric evaporation is about 
tens of $\mathrm{km\ s^{-1}}$ (Li et al. 2019) that is far below the $V_{\mathrm{max}}$ in GWAC\,201221A, and far below both bulk velocity and 
$V_{\mathrm{max}}$ in GWAC\,210117A. Although a chromospheric evaporation with $V_{\mathrm{max}}\sim600\ \mathrm{km\ s^{-1}}$ 
was reported by Gunn et al. (1994) in a dMe4.5 star, the possibility of association with a CME is argued recently by Koller et al. (2021).
The observations on the Sun indicate that the red asymmetry in H$\alpha$ emission line due to either chromospheric condensation or coronal rain shows velocity no more than 
$\sim100\ \mathrm{km\ s^{-1}}$ (e.g., Asai et al. 2012; Ichimoto \& Kurokawa 1984), which is incompatible with the large bulk and (or)
$V_{\mathrm{max}}$ observed in the current two flares.

\rm
\begin{figure}[ht!]
\plotone{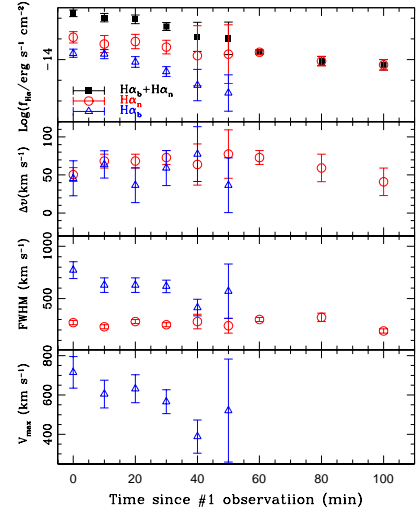}
\caption{In GWAC\,201221A, the temporal evolution of H$\alpha$ line flux, bulk velocity, line width and maximum projected velocity ($V_{\mathrm{max}}$) during the flare are shown in the panels in an order from top to bottom.
\label{fig:general}}
\end{figure}

\begin{figure}[ht!]
\plotone{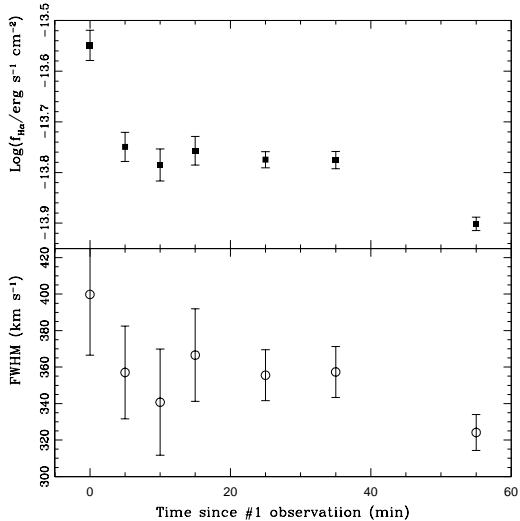}
\caption{\it Upper panel: \rm Temporal evolution of H$\alpha$ line flux for GWAC\,210117A. 
\it Lower panel: \rm The same as the upper one but for H$\alpha$ line width.
\label{fig:general}}
\end{figure}

\subsection{CME Mass}
The total line luminosity caused by a transition from level $j$ to $i$ is $L_{ji}=N_jA_{ji}h\nu_{ji}VP_{\mathrm{esc}}$,
where $A_{ji}$ is the Einstein coefficient for a spontaneous decay from level $j$ to $i$,
$N_j$ the number density of hydrogen atoms at excited level $j$, $V$ the total volume and 
$P_{\mathrm{esc}}$ the escape probability. 
By considering $M_{\mathrm{CME}}\geq N_{\mathrm{tot}}Vm_{\mathrm{H}}$, the total hydrogen mass involved in a CME can then be 
estimated as (Koller et al. 2021; Houdebine et al. 1990):
\begin{equation}
M_{\mathrm{CME}}\geq\frac{4\pi d^2f_{\mathrm{line}} m_{\mathrm{H}}}{h\nu_{ji}A_{ji}P_{\mathrm{esc}}}\frac{N_{\mathrm{tot}}}{N_j}
\end{equation}
where $d$ is the distance to the star, $f_{\mathrm{line}}$ the line flux due to a CME, $m_{\mathrm{H}}$ the mass of the hydrogen atom and $N_{\mathrm{tot}}$ is the number density of natural hydrogen atoms.

After transforming the observed H$\alpha$ line flux to H$\gamma$ by adopting a Balmer decrement of 3 (Butler et al. 1988), 
a mass of $1.5-4.5\times10^{19}$g and $7.1\times10^{18}$g can be obtained for GWAC\,201221A and GWAC\,210117A, respectively.
In the calculations, we adopt
$A_{52}=2.53\times10^6$ (Wiese \& Fuhr 2009) and $N_{\mathrm{tot}}/N_5=2\times10^9$ estimated from nonlocal thermal equilibrium modeling (Houdebine \& Doyle 1994a, b)\footnote{We estimate the CME mass from the H$\gamma$ emission line, instead of H$\alpha$, because of a lack of $N_3/N_{\mathrm{tot}}$ in literature.}. 
We emphasize that only the modeled broad H$\alpha$ components are used in the 
mass estimations. 
A typical value of 0.5 is adopted for $P_{\mathrm{esc}}$ in the calculations (Leitzinger et al. 2014).   
We argue that the estimated masses are in fact 
in agreement with the previously expected range of $10^{14-19}$g of a stellar CME (Moschou et al. 2019).

\subsection{Flare energy}

On the one hand, based on the method described in Kowalski et al. (2013), the total energy $E_R$ in the $R-$band can
be estimated from the formula: $E_R= 4\pi d^2\times f_{R,\mathrm{q}}\times ED$, 
where $f_{R,\mathrm{q}}$ is the quiescent $R-$band flux determined previously in Section 2.
With the distance and the determined 
ED, $E_R$ is resulted to be 
$\sim1.6\times10^{35}\ \mathrm{erg}$ and $\sim8.3\times10^{33}\ \mathrm{erg}$ 
for GWAC\,201221A and GWAC\,210117A, respectively. The bolometric energy can then be estimated to 
be $E_{\mathrm{bol}}\sim9.6\times10^{35}\ \mathrm{erg}$ and $\sim5.0\times10^{34}\ \mathrm{erg}$ by adopting a bolometric correction of $L_{\mathrm{bol}}/L_{R}=6$.
The bolometric correction is obtained by
assuming a blackbody with an effective temperature of $T_{\mathrm{eff}}=10^4$K. 

On the other hand, it is known that there is a relationship between flare X-ray luminosity and H$\alpha$ 
luminosity given the statistical studies in both solar-like stars and more active stars 
(e.g., Bulter 1993; Martinez-Arnaiz et al. 2011). Based on the simple linear relationship of
$L_{\mathrm{X}}=16L_{\mathrm{H\alpha}}$ (i.e., Equation (1) in Mouschou et al. (2019)) and our spectroscopic and light curve analysis, 
the X-ray fluence of the flare is estimated to be 
$F_{\mathrm{X}}=5.3\times10^{34}\ \mathrm{erg}$ and $2.1\times10^{34}\ \mathrm{erg}$ 
for GWAC\,201221A and GWAC\,210117A, respectively.

The X-ray luminosity can be alternatively estimated from the H$\alpha$ emission according to 
the relationship, involving stellar surface flux, provided by Martinez-Arnaiz et al. (2011):
\begin{equation}
  \log L_\mathrm{X}=-2.72+1.48\log L_{\mathrm{H\alpha}}-0.92\log R_*
\end{equation}
where $R_*$ is the stellar radius. The relationship returns an X-ray fluence of 
$F_{\mathrm{X}}=8.4\times10^{34}\ \mathrm{erg}$ and $2.7\times10^{34}\ \mathrm{erg}$ for GWAC\,201221A and GWAC\,210117A, respectively, which are only slightly larger than the above values based on the simple 
linear relationship. 

The ratio of white light to X-ray components in a stellar flare is now still poorly understood.
With the above estimations, the ratio of $F_{\mathrm{X}}/E_{\mathrm{bol}}$ is $\approx0.05-0.08$ in 
GWAC\,201221A, which roughly agrees with the result of $\approx0.01$ revealed by the solar observations 
(e.g., Kretzschmar 2011; Emslie et al. 2012). A much larger $F_{\mathrm{X}}/E_{\mathrm{bol}}\approx0.4-0.5$ is, however, obtained for GWAC\,210117A, which could be explained by the fact that 
Martinez-Arnaiz et al. (2011) suggests that there is an equipartition between X-ray and Balmer line emission for the flares with low fluence.


\section{Conclusion}

Thanks to the high cadence survey of the GWAC system, 
fast and time-resolved spectroscopic follow-ups are carried out for two M-dwarf flares. 
The light curve analysis indicates a flare energy in $R-$band of $1.6\times10^{35}\ \mathrm{erg}$ and $8.1\times10^{33}\ \mathrm{erg}$. High velocity wings of their H$\alpha$ emission lines can be 
identified in both flares. The large projected maximum velocity of $\sim500-700\ \mathrm{km\ s^{-1}}$ motivates us to argue that the wings are most likely resulted from a CME event, 
after excluding the possibility of chromospheric evaporation
and coronal rain. With the modeling of the H$\alpha$ emission line profiles, 
the CME masses are estimated to be  $1.5-4.5\times10^{19}\ \mathrm{g}$ and $7.1\times10^{18}\ \mathrm{g}$.

\acknowledgments


The authors thank the anonymous referee for a careful review
and helpful suggestions that improved the manuscript greatly.
The authors are thankful for support from the National Key R\&D Program
of China (grant No. 2020YFE0202100). The study is supported
by the National Natural Science Foundation of China under
grants 11773036 and U1821207, and by the Strategic Pioneer Program on
Space Science, Chinese Academy of Sciences, grants No.
XDA15052600 and XDA15016500. J.W. is supported by the
Natural Science Foundation of Guangxi (2020GXNSFDA238018),
and by the Bagui Young Scholars Program. 
We thank the night assistants and duty astronomers of the NAOC 2.16m telescope and 
the GWAC system for their instrumental and observational help.
This work has made use of data from the European
Space Agency (ESA) mission Gaia (https://www.cosmos.esa.int/gaia), 
processed by the Gaia Data Processing and Analysis
Consortium (DPAC, https://www.cosmos.esa.int/web/gaia/dpac/consortium). 
Funding for the DPAC has been provided
by national institutions, in particular the institutions 
participating in the Gaia Multilateral Agreement. This research has made
use of the VizieR catalog access tool, CDS, Strasbourg, France
(doi: 10.26093/cds/vizier). The original description of the
VizieR service was published in A\&AS 143, 23.

\vspace{5mm}
\facilities{GWAC, GWAC-F60A, NAOC 2.16m telescope}
\software{IRAF (Tody 1986, 1992), Python}


\vskip12pt


\begin{thebibliography}{}
\bibitem[Aarnio et al. (2011)]{aar11} Aarnio, A. N., Stassun, K. G., Hughes, W. J., \& McGregor, S. L. 2011, \solphys, 268, 195 
\bibitem[Ambruster et al. (1989)]{amb89}  Ambruster, C. W., Pettersen, B. R., \& Sundland, S. R. 1989, \aap, 208, 198  
\bibitem[Anders et al. (2019)]{and94} Anders, F., Khalatyan, A., Chiappini, C., Queiroz, A. B., Santiago, B. X., Jordi, C., Girardi, L., Brown, A. G. A., et al. 2019, \aap, 628, 94
\bibitem[Antolin et al. (2012)]{ant12} Antolin, P., Vissers, G., \& Rouppe van der Voort, L. 2012,  \solphys, 280, 457 
\bibitem[Argiroffi et al. (2019)]{arg19} Argiroffi, C., Reale, F., Drake, J. J., et al. 2019, Nature Astronomy, 3, 742
\bibitem[Asai et al. (2019)]{asa19} Asai, A., Ichimoto, K., Kita, R., Kurokawa, H., \& Shibata, K. 2019, \pasj, 64, 20
\bibitem[Balona (2015)]{bal15} Balona, L. A. 2015, \mnras, 447, 2714 
\bibitem[Berdyugina et al. (1999)]{ber99} Berdyugina, S. V., Ilyin, I., \& Tuominen, I. 1999, \aap, 349, 863
\bibitem[Boudreault et al. (2012)]{bou12} Boudreault, S., Lodieu, N., Deacon, N. R., \& Hambly, N. C. 2012, \mnras, 426, 3419 
\bibitem[Butler (1993)]{but93} Butler, C. J. 1993, \aap, 272, 507
\bibitem[Butler et al. (1988)]{but88} Butler, C. J., Rodono, M., \& Foing, B. H. 1988, \aap, 206, L1
\bibitem[Cantat-Gaudin et al. (2020)]{can20} Cantat-Gaudin, T., Anders, F., Castro-Ginard, A., Jordi, C., Romero-Gomez, M., Soubiran, C., Casamiquela, L., Tarricq, Y., et al. 2020, \aap, 640, 1 
\bibitem[Canfield et al. (1990)]{can90} Canfield, R. C., Penn, M. J., Wulser, J., \& Kiplinger, A. L. 1990, \apj, 363, 318
\bibitem[Chandra et al. (2016)]{cha16} Chandra, R., Chen, P. F., Fulara, A., Srivastava, A. K., \& Uddin, W. 2016, \apj, 822, 106
\bibitem[Chang et al. (2018)]{cha18} Chang, H. -Y., Lin, C. -L., Ip, W. -H., Huang, L. -C., Hou, W. -C., Yu, P. -C., Song, Y. -H., \& Luo, A. 2018, \apj, 867, 78
\bibitem[Cherenkov et al. (2017)]{che17} Cherenkov, A., Bisikalo, D., Fossati, L., \& Mostl, C. 2017,  \apj, 846, 31 
\bibitem[Davenport et al. (2014)]{dav14} Davenport, J. R. A., Hawley, S. L., Hebb, L., et al. 2014, \apj, 797, 122
\bibitem[Davenport et al. (2016)]{dav16} Davenport, J. R. A., Kipping, D. M., Sasselov, D., Matthews, J. M., \& Cameron, C. 2016, \apjl, 829, 31
\bibitem[Emslie et al. (2012)]{ems12} Emslie, A. G., Dennis, B. R., Shih, A. Y., et al. 2012, \apj, 759, 71
\bibitem[Fan et al. (2016)]{fan16} Fan, Z., Wang, H. J., Jiang, X. J., et al. 2016, \pasp, 128, 5005
\bibitem[Forbes et al. (2006)]{for06} Forbes, T. G., Linker, J. A., Chen, J., et al. 2006, SSR,v, 123, 251  
\bibitem[Fuhrmeister et al. (2018)]{fuh18} Fuhrmeister, B., Czesla, S., Schmitt, J. H. M. M., et al. 2018, \aap, 615, 14  
\bibitem[Gaia Collaboration et al. (2018)]{gai18} Gaia Collaboration, Brown, A. G. A., Vallenari, A., et al. 2018a, \aap, 616, 1
\bibitem[Gaia Collaboration et al. (2018)]{gai18} Gaia Collaboration, Brown, A. G. A., Vallenari, A., et al. 2018b, \aap, 616, 10
\bibitem[Gunn et al. (1994)]{gun94} Gunn, A. G., Doyle, J. G., Mathioudakis, M., Houdebine, E. R., \& Avgoloupis, S. 1994, \aap, 285, 489
\bibitem[Hardegree-Ullman et al. (2020)]{har20} Hardegree-Ullman, K. K.,  Zink, J. K., Christiansen, J. L., Dressing, C. D., Ciardi, D. R., \& Schlieder, J. E. 2020, \apjs, 247, 28
\bibitem[Houdebine \& Doyle (1994a)]{hod94a}  Houdebine, E. R., \& Doyle, J. G. 1994a, \aap, 289. 169
\bibitem[Houdebine \& Doyle (1994b)]{hod94b}  Houdebine, E. R., \& Doyle, J. G. 1994b, \aap, 289, 185
\bibitem[Houdebine ert al. (1990)]{hou90} Houdebine, E. R., Foing, B. H., \& Rodono, M. 1990, \aap, 238, 249 
\bibitem[Huenemoerder et al. (2010)]{hue10} Huenemoerder, D. P., Schulz, N. S., Testa, P., Drake, J. J., Osten, R. A., \& Reale, F. 2010, \apj, 723, 1558 
\bibitem[Ichimoto \& Kurokawa (1984)]{ick84} Ichimoto, K., \& Kurokawa, H. 1984, \solphys, 93, 105
\bibitem[Ilin et al. (2019)]{ili19} Ilin, E., Schmidt, S. J., Davenport, J. R. A., \& Strassmeier, K. G. 2019, \aap, 622, 133 
\bibitem[Karlicky \& Barta (2007)]{kab07} Karlicky, M., \& Barta, M. 2007, \aap, 464, 735
\bibitem[Kliem et al. (2000)]{kli00} Kliem, B., Karlicky, M., \& Benz, A. O. 2000, \aap, 360,715
\bibitem[Koller et al. (2021)]{kol21} Koller, F., Leitzinger, M., Temmer, M., Odert, P., Beck, P. G., \& Veronig, A. 2021, \aap, 646, 34
\bibitem[Kowalski et al. (2013)]{kow13} Kowalski, A. F., Hawley, S. L., Wisniewski, J. P., Osten, R. A., Hilton, E. J., Holtzman, J. A., Schmidt, S. J., Davenport, J. R. A. 2013, \apjs, 207, 15  
\bibitem[Kretzschmar (2011)]{kre11} Kretzschmar, M. 2011, \aap, 530, 84
\bibitem[Kriss (1994)]{kri94} Kriss, G. 1994, in ASP Conf. Ser. 61, Astronomical Data Analysis Software
and Systems III, ed. D. R. Crabtree, R. J. Hanisch, \& J. Barnes (San Fransisco, CA: ASP), 437
\bibitem[Lacatus et al. (2012)]{lac12} Lacatus, D. A., Judge, P. G., Donea, A., Daniela A., Judge, P. G., \& Donea, A. 2012, \apjl, 842, 15 
\bibitem[Leitzinger et al. (2014)]{lei14} Leitzinger, M., Odert, P., Greimei, R., Korhonen, H., Guenther, E. W., Hanslmeier, A., Lammer, H., \& Khodachenko, M. L. 2014, \mnras, 443, L898 
\bibitem[Li et al. (2019)]{lid19} Li, Y., Ding, M. D., Hong, J., Li, H., \& Gan, W. Q. 2019, \apj, 879, 30
\bibitem[Li et al. (2016)]{li16} Li, L., Zhang, J., Peter, H., et al. 2016, Nature Physics, 12,847
\bibitem[Martinez-Arnaiz et al. (2011)]{mar11} Martinez-Arnaiz, R., Lopez-Santiago, J., Crespo-Chacon, I., \& Montes, D. 2011, \mnras, 414, 2629
\bibitem[Massey et al. (1988)]{mas88} Massey, P., Strobel, K., Barnes, J. V., et al. 1988, \apj, 328, 315
\bibitem[Moschou et al. (2017)]{mos17} Moschou, S., Drake, J. J., Cohen, O., Alvarado-Gomez, J. D., \& Garraffo, C. 2017, \apj, 850, 191
\bibitem[Moschou et al. (2019)]{mos19} Moschou, S., Drake, J. J., Cohen, O., Alvarado-Gomez, J. D., Garraffo, C., \& Fraschetti, F. 2019, \apj, 877, 105
\bibitem[Notsu et al. (2016)]{not16} Notsu, Y., Maehara, H., Shibayama, T., Honda, S., Notsu, S., Namekata, K., Nogami, D., \& Shibata, K. 2016, The 19th Cambridge Workshop on Cool Stars, Stellar Systems, and the Sun (CS19), Uppsala, Sweden, 06-10 June 2016, id.119
\bibitem[Noyes e tal. (1984)]{noy84} Noyes, R. W., Hartmann, L. W., Baliunas, S. L., Duncan, D. K., \& Vaughan, A. H. 1984, \apj, 279, 763 
\bibitem[Odert et al. (2020)]{ode20} Odert, P., Leitzinger, M., Guenther, E. W., \& Heinzel, P. 2020, \mnras, 494, 3766 
\bibitem[Osten et al. (2004)]{ost04} Osten, R. A., Brown, A., Ayres, T. R., et al. 2004, \apjs, 153, 317 
\bibitem[Osten et al. (2005)]{ost05} Osten, R. A., Hawley, S. L., Allred, J. C., Johns-Krull, C. M., \& Roark, C. 2005, \apj, 621, 398
\bibitem[Paudel et al. (2018)]{pau18} Paudel, R. R., Gizis, J. E., Mullan, D. J., Schmidt, S. J., Burgasser, A. J., Williams, P. K. G., \& Berger, E. 2018, \apj, 858, 55
\bibitem[Pecaut \& Mamajek (2013)]{pem13} Pecaut, M. J., \& Mamajek, E. E. 2013, \apjs, 208, 9 
\bibitem[Pettersen (1989)]{pet89} Pettersen, B. R.  1989, \aap, 209, 279
\bibitem[Perez-Montero \& Diaz (2013)]{ped13} Perez-Montero, E., \& Diaz, A. I. 2013, \mnras, 346, 105 
\bibitem[Schlafly \& Finkbeiner (2011)]{sch11} Schlafly, E. F., \&  Finkbeiner, D. P.  2011, \apj, 737, 103
\bibitem[Schmidt et al. (2019)]{sch19} Schmidt, S. J., Shappee, B. J., van Saders, J. L.. et al. 2019, \apj, 876, 115
\bibitem[Schmitt (1994)]{sch94} Schmitt, J. H. M. M. 1994, \apjs, 90, 735
\bibitem[Shulyak et al. (2017)]{shu17} Shulyak, D., Reiners, A., Engeln, A., Malo, L., Yadav, R., Morin, J., \& Kochukhov, O. 2017, Nature Astronomy, 1, 184
\bibitem[Stassun et al. (2019)]{sta19} Stassun, K. G., Oelkers, R. J., Paegert, M., Torres, G., Pepper, J., De Lee, N., Collins, K., Latham, D. W., et al., 2019, \aj, 158, 138
\bibitem[Tody (1986)]{tod86} Tody, D. 1986, Proc. SPIE, 627, 733
\bibitem[Tody (1993)]{tod93} Tody, D. 1993, in ASP Conf. Ser. 52, Astronomical Data Analysis Software and Systems II, ed. R. J. Hanisch, R. J. V. Brissenden, \& J. Barnes (San Fransisco, CA: ASP), 173
\bibitem[Tsuneta (1996)]{tsu96} Tsuneta, S. 1996, \apj, 456, 840
\bibitem[Van Doorsselaere et al. (2017)]{van17} Van Doorsselaere, T., Shariati, H., \& Debosscher, J. 2017, \apjs, 232, 26 
\bibitem[Vida et al. (2019)]{vid19} Vida, K., Leitzinger, M., Kriskovics, L., Seli, B., Odert, P., Kovacs, O., Korhonen, H., van Driel-Gesztelyi, L. 2019, \aap, 623, 49
\bibitem[Wang et al. (2020)]{wan20} Wang, J., Li, H. L., Xin, L. P., et al. 2020, \aj, 159, 35
\bibitem[Wei et al. (2016)]{wei16} Wei, J. Y., Cordier, B., Antier, S., et al. 2016, arXiv:1610.0689
\bibitem[Webb \& Howard (2012)]{weh12} Webb, D. F., \& Howard, T. A. 2012, LRSP, 9, 3
\bibitem[Wiese \& Fuhr (2009)]{wif09} Wiese, W. L., \& Fuhr, J. R. 2009, JPCRD, 38, 565
\bibitem[Wright et al. (2011)]{wri11} Wright, N. J., Drake, J. J., Mamajek, E. E., \& Henry, G. W. 2011, \apj, 743, 48
\bibitem[Xin et al. (2020)]{xin20} Xin, L. P., Li, H. L., Wang, J., et al. 2021, \apj, 909, 106  
\bibitem[Xu et al. (2020)]{xu20}  Xu, Y., Xin, L. P., Wang, J., Han, X. H., Qiu, Y. L., Huang, M. H., \& Wei, J. Y. 2020, \pasp, 132, 054502
\bibitem[Yashiro et al. (2008)]{yas08} Yashiro, S., Michalek, G., Akiyama, S., Gopalswamy, N., \& Howard, R. A. 2008, \apj, 673, 1174 

 

\end{thebibliography}
\end{document}